\documentclass{article}
\usepackage{spconf,amsmath,graphicx}
\usepackage{hyperref}
\usepackage{url}
\usepackage{tikz}
\usepackage{color, colortbl}
\usepackage{subcaption}

\title{Mellotron: Multispeaker expressive voice synthesis by conditioning on rhythm, pitch and global style tokens}

\name{Rafael Valle*, Jason Li*, Ryan Prenger, Bryan Catanzaro}
\address{NVIDIA Corporation}

\begin{document}

\maketitle

\begin{abstract}
%auto-ignore
Mellotron is a multispeaker voice synthesis model based on Tacotron 2 GST that can make a voice emote and sing without emotive or singing training data. By explicitly conditioning on rhythm and continuous pitch contours from an audio signal or music score, Mellotron is able to generate speech in a variety of styles ranging from read speech to expressive speech, from slow drawls to rap and from monotonous voice to singing voice. Unlike other methods, we train Mellotron using only read speech data without alignments between text and audio. We evaluate our models using the LJSpeech and LibriTTS datasets. We provide F0 Frame Errors and synthesized samples that include style transfer from other speakers, singers and styles not seen during training, procedural manipulation of rhythm and pitch and choir synthesis.  
\end{abstract}


\section{Introduction}\label{sec:introduction}
%auto-ignore
% Explain to the reader what the problem is
Speech synthesis is typically formulated as the conversion of text to speech (TTS). This formulation, however, leaves out control for all the aspects of speech not contained in the text. Here we approach the problem of expressive speech synthesis which includes not just text, but other characteristics such as pitch, rhythm and emphasis. There are formulations to expressive speech synthesis that require animated and emotive voice data. This is an inconvenient drawback given the limited access to such data. In our approach, we can make a voice emote and sing without any such data.

% Explain to the reader where others fail and we succeed!
%(rprenger) tripping over definition of expressive speech here.  It must include singing, but "speech" is in the title
Recent approaches that utilize deep learning for expressive speech synthesis combine text and a learned latent embedding for prosody or global style \cite{skerry2018towards,wang2018style}. While these approaches have shown promise, manipulating such latent variables only offers a coarse control over expressive characteristics of speech. Mellotron was motivated by the desire for fine grained control over these expressive characteristics. Notably, we show that it is easy to condition Mellotron on pitch and rhythm information automatically extracted from an audio signal or music score.

By accounting for melodic information such as pitch and rhythm, expressive speech synthesis with Mellotron can be easily extended to singing voice synthesis (SVS) \cite{Nishimura2016SVS,lee2019adversarially}. Unfortunately, recent attempts \cite{lee2019adversarially} require a singing voice dataset and heavily quantized pitch and rhythm data obtained from a digital representation of a music score, for example MIDI \cite{moog1986midi} or musicXML \cite{good2001musicxml}. Mellotron does not require any singing voice in the dataset nor manually aligned pitch and text in order to synthesize singing voice.

%We extend previous approaches \cite{skerry2018towards,wang2018style,Nishimura2016SVS,lee2019adversarially}, by conditioning mel-spectrogram generation on explicit and latent variables, hence combining their power. This factorization allows for easy procedural manipulation of a speaker's style by controlling their rhythm and pitch. Mellotron can be trained from common read speech datasets without the need for hand alignments between text and audio nor between pitch and audio. Furthermore, although Mellotron does not require training on singing voice data, it is able to generalize to singing voice synthesis.
%

Mellotron can make a voice emote and sing without emotion or singing data. Training Mellotron is very simple and only requires read speech and transcriptions. During inference, we can change the generated voice's speaking style, make it emote or sing by extracting pitch and rhythm characteristics from an audio file or a music score. As a bonus, with Mellotron we can explore latent characteristics from an audio corpus by sampling a dictionary of learned latent characteristics. In summary, Mellotron is a versatile voice synthesis\footnote{Includes speech synthesis, singing voice synthesis, etc.} model that enables the combination of characteristics from different sources and generalizes to characteristics not seen in training data.
\section{Method}\label{sec:methods}
%auto-ignore
%(rprenger) A lot of what's in this first paragraph could be moved to the intro  We don't need to compare to other methods in our Methods section
Mellotron is a voice synthesis model that uses a combination of explicit and latent variables. Whereas well-established signal processing algorithms provide explicit variables that are valuable to expressive speech such as fundamental frequency contours and voicing decisions, deep learning strategies can be used to learn latent variables that express characteristics of an audio corpus that are unknown to the user and hard to formalize.

We factorize a single speaker mel-spectrogram $M$ into explicit variables such as text, speaker identity, a fundamental frequency contour augmented with voiced/unvoiced decisions and two latent variables learned by the model during training. The first latent variable refers to a dictionary of vectors that can be queried with an audio input or sampled directly as described in \cite{wang2018style}. The second latent variable is the learned attention map between the text and the mel-spectrogram as described in \cite{shen2018natural}. 

From now on we will refer to the augmented fundamental frequency contour as pitch contour and refer to the first and second latent variables as global style tokens (GST) and rhythm respectively.

We are interested in factorizing $M = [T, S, P, R, Z]$, where T represents the text, S represents the speaker identity, P represents the pitch contour, R represents the rhythm and Z represents the global style tokens. Given this formulation, during training we maximize the following:

\begin{equation}
    P(mel^{(i)}|T^{(i)}, S^{(i)}, P^{(i)}, R^{(i)}, Z_{mel^{(i)}};\theta),
    \label{eq:mellotron}
\end{equation}

where the superscript $i$ represents the i-th mel, $T^{(i)}$, $S^{(i)}$ and $P^{(i)}$ represent the text, speaker, and pitch contour associated with the i-th mel, $R^{i}$ represents the learned alignments between the text and mel-spectrogram frames, $Z_{mel^{(i)}}$ represents the global style token conditioned on $mel^{(i)}$ as presented in \cite{wang2018style}, and $\theta$ represents the model parameters.

The explicit factors offers two advantages. First, by providing the model with text and speaker information, we prevent the problem of entanglement between text and speaker information. Second by providing the model with pitch contour and voicing information, we are able to directly control pitch and voicing decisions during inference.

Similarly the latent factors offers two advantages. First, by learning the alignment map between the text and mel-spectrogram during training, we do not need to extract phoneme alignments for training and can control the rhythm during inference by providing the model with an alignment map. Second, by providing the model with a dictionary of latent variables, we are able to learn latent factors that are harder to express or extract explicitly, thus leveraging the full power of latent variables.

\label{sec:implementation_inference}
Using this formulation we are able to transfer the text, rhythm and pitch contour from a source, e. g. audio signal or musical score, to a target speaker by replacing the variables in Equation \ref{eq:mellotron} accordingly. For example, we first collect the text, pitch and rhythm ($T_s, P_s, R_s$), from the source, sample a GST $Z_{query}$ from the GST dictionary learned by Mellotron, and chose a target speaker $S_t$.

\begin{equation}
    P(mel_{out}|T_s, P_s, S_t, R_s, Z_{query};\theta)
\end{equation}

$mel_{out}$ should now have the same text, pitch and rhythm as the source, latent characteristics obtained from the global style token and the voice of the target speaker. In our current formulation, the target speaker, $S_t$, would always be found in the training set, while the source text, pitch and rhythm ($T_s, P_s, R_s$) could be from outside the training set. This allows us to train a model that makes a voice emote and sing without using any singing voice in the training dataset, without any manual labelling of emotions nor pitch, and without any manual alignments between words and audio, nor between pitch and audio.
\section{Implementation}\label{sec:implementation}
%auto-ignore
In this section we are going to describe our model architecture and our training and inference setups. We plan to release our implementation and pre-trained models on github.

\subsection{Architecture}
Mellotron extends Tacotron 2 GST \cite{wang2018style} with speaker embeddings and pitch countours. Unlike \cite{gibiansky2017deep, ping2017deep}, where site specific speaker embeddings are used, we use a single speaker embedding that is channel-wise concatenated with the encoder outputs over every token. The pitch contour goes through a single convolution layer followed by a ReLU non-linearity. We experiment with kernel sizes 1 and 3 and convolution dimensions 1 and 8. The pitch contour is channel-wise concatenated with the decoder inputs. 
We use phoneme representations whenever possible.

\subsection{Training}
Our implementation only requires text and audio pairs with a speaker id. Our pitch contours are automatically extracted using the Yin algorithm \cite{de2002yin} with harmonicity thresholds between 0.1 and 0.25. Unlike \cite{lee2019adversarially}, during training our model does not require manually aligned text, pitch and mel-spectrogram.  We use the L2 loss between ground truth and predicted mels described in \cite{wang2018style} without any modifications.

\subsection{Inference}
Following the description in Section \ref{sec:implementation_inference}, during inference we provide  Melloron with text, rhythm and pitch information that is obtained either from an audio signal or from a musical score, a global style token and a speaker id.
% During our qualitative experiments we find that some of the pitch contours seem to be outside of a speaker's vocal range. When this happens, Mellotron defaults to a constant highest or lowest pitch value. We circumvent this by scaling the pitch contour by a constant such that it matches the speaker's vocal range.

\subsubsection{Audio Signal}
Obtaining text, rhythm and pitch information consists of three steps. First, we extract text information from an audio file by either using an automatic speech recognition model \cite{li2019jasper,chiu2018state} or by manually transcribing the text. The text information is pre-processed with our text cleaners and then converted from graphemes to phonemes.

Second, we extract rhythm information by using a forced-alignment tool \cite{gentle, mcauliffe2017montreal} or by using Mellotron as a forced-aligner. Alignment maps can be obtained with Mellotron by performing a teacher-forced forward pass using the data from the source signal. Whenever necessary, we fine tune the alignment maps by hand or by training Mellotron on the source signal for a few iterations with small learning rate.

The pitch data is obtained by using Yin \cite{de2002yin} or Melodia \cite{salamon2012melody}. In our quantitative experiments we use Yin to replicate the setup described in \cite{skerry2018towards}. In our qualitative experiments we use Melodia instead as we find it to be more precise than Yin, specially with regards to false voiced decisions. 

\subsubsection{Music Score}
We operate on music scores in XML format containing event tuples with pitch, note duration and syllables for each part in the score. We directly convert pitch to frequency and use the FFT hop size to convert event durations from seconds to frames. We remind the reader that although we refer to pitch, our model's representation of pitch is continuous.

We concatenate the syllables into words and convert graphemes to phonemes. For single phone events, the duration of each phone is equal to the duration of the event. For multi-phone events, the duration of each phone is dependent on its type: we use heuristics to assign durations between 20 and 100ms to consonants and assign the remainder of the event's duration to vowels. For example, consider a one second long single note event on the word \textit{Bass} with phoneme representation [B, AE, S]. We set B to 20 ms, S to 100 ms and the remaining duration to AE, and hence have full control over the duration of each phone.
\section{Experiments}\label{sec:experiments}
%auto-ignore
We train our models using the LJSpeech (LJS) dataset \cite{ito2017lj}, the Sally dataset, a proprietary single speaker dataset with 20 hours, and a subset of LibriTTS \cite{zen2019libritts}. All datasets used in our experiments are from read speech.

We provide results that include style transfer\footnote{Transferring text, rhythm and pitch contour to a target speaker.} from source speakers seen and unseen in the dataset, from singers, procedural manipulation of rhythm and choir synthesis from music scores. Visit our \href{https://nv-adlr.github.io/Mellotron}{website}\footnote{\url{https://nv-adlr.github.io/Mellotron}} to listen to Mellotron samples.

\subsection{Training Setup}
For all the experiments, we trained on LJS, Sally and the \textit{train-clean-100} subset of LibriTTS with over 100 speakers and 25 minutes on average per speaker. Speakers with less than 5 minutes of data and files that are larger than 10 seconds were filtered out.  We do not perform any data augmentation, hence any extension to a speaker's characteristics such as vocal range and speech rate is made possible with Mellotron.

We use a sampling rate of 22050 Hz and mel-spectrograms with 80 bins using librosa mel filter defaults. We apply the STFT with a FFT size of 1024, hop size of  256, and window size of 1024 samples.

We use the ADAM \cite{kingma2014adam} optimizer with default parameters, start with a 1e-3 learning rate and anneal the learning rate as the loss starts to plateau. We decrease training time by using a single NVIDIA DGX-1 with 8 GPUs.

For decoding the mel-spectrograms produced by Mellotron, we use a single WaveGlow \cite{prenger2019waveglow} model trained on the Sally dataset. Our results suggest that Waveglow can be used as an universal decoder.

In our setup, we find it easier to first learn attention alignments on speakers with large amounts of data and then fine tune to speakers with less data. Thus, we first train Mellotron on LJS and Sally and finetune it with a new speaker embedding on LibriTTS, starting with a learning rate of 5e-4 and annealing the learning rate as the loss starts to plateau.

\label{sec:quantitative_results}
\subsection{Quantitative Results}
In this section we provide quantitative results that compare Gross Pitch Error (GPE) \cite{Nakatani2008}, Voicing Decision Error (VDE) \cite{Nakatani2008} and F0 Frame Error (FFE) \cite{chu2009reducing} between Mellotron and E2E-Prosody \cite{skerry2018towards}. 
Following \cite{skerry2018towards}, all pitch and voicing metrics are computed using the Yin algorithm \cite{de2002yin}. Due to the rhythm conditioning, our reference and predicted audio have the same length and does not require padding.

The results in Table~\ref{tab:ffe} below show that by conditioning on pitch we can drastically reduce the error between the source and the synthesized voice. For singing voice, low pitch error is extremely important otherwise the melody might lose its identity. For prosody transfer, a lower FFE provides evidence that the style will be more precisely transferred to the target.
\renewcommand{\arraystretch}{1.1}
\begin{table}[!ht]
\begin{center}
    \begin{tabular}{ c|c|c|c|c}
        \textbf{Model} & \textbf{Voice} &  \textbf{GPE}& \textbf{VDE}& \textbf{FFE}\\
        \hline
        E2E-Prosody    & Single         & - & - & $28.1\%$\\
        E2E-Prosody    & Multi          & - & - & $27.5\%$\\
        Mellotron      & LJS-Sally      & 0.08\% & 9.19\% & $9.28\%$\\
        Mellotron      & LibriTTS       & 0.08\% & 8.69\% & $8.77\%$\\
        \hline
    \end{tabular}
    \caption{GPE, VDE, FFE for Mellotron and E2E-Prosody. The reference is always the same speaker.}
    \label{tab:ffe}
\end{center}
\end{table}

\subsection{Style transfer from Audio Signal}
Mellotron is able to emote and match the style of an input audio by replicating its rhythm or both its rhythm and pitch. Overall, we note that our experiments using audio data are directly impacted by the quality of the rhythm and pitch contours provided to the model. Whereas Melodia provides rather precise pitch contours, we find that the rhythm data obtained from forced-alignments had to be constantly fine-tuned. In all audio experiments we obtain the rhythm by fine-tuning alignment maps obtained by using Mellotron as a forced-aligner. Occasionally we find that some of the pitch contours seem to be outside of a speaker's vocal range. When this happens, Mellotron defaults to a constant highest or lowest pitch value. We circumvent this by scaling the pitch contour by a constant to matches the speaker's vocal range.

\subsubsection{Rhythm Transfer}
In this experiment we transfer the rhythm and its associated text from a source audio signal to a target speaker. Our formulation provides procedural control over the duration of every phoneme, hence allowing for simple manipulations such as changing the speech rate or complex effects like speeding up or slowing down. In rhythm transfer, we provide Mellotron with an array of zeros as the pitch contour. 

\begin{figure}[!ht]
    \centering
    \begin{subfigure}{0.49\linewidth}
        \centering
        \includegraphics[width=\textwidth]{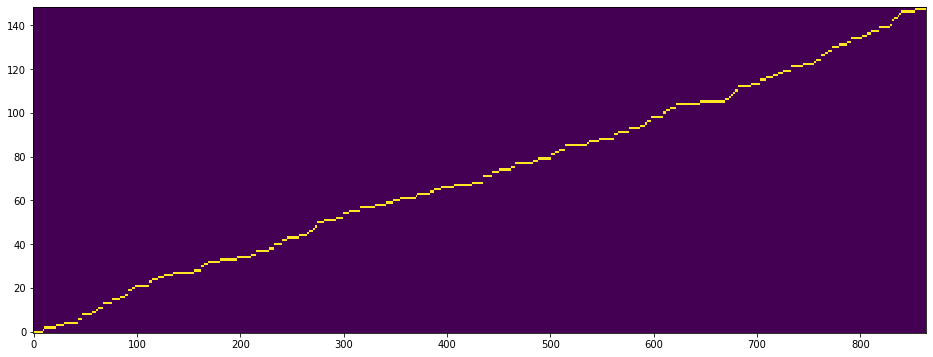}
        %\caption{Source alignment}
        %\label{fig:reference_alignment}
    \end{subfigure}
    \hfill
    \begin{subfigure}{0.49\linewidth}
        \centering
        \includegraphics[width=\textwidth]{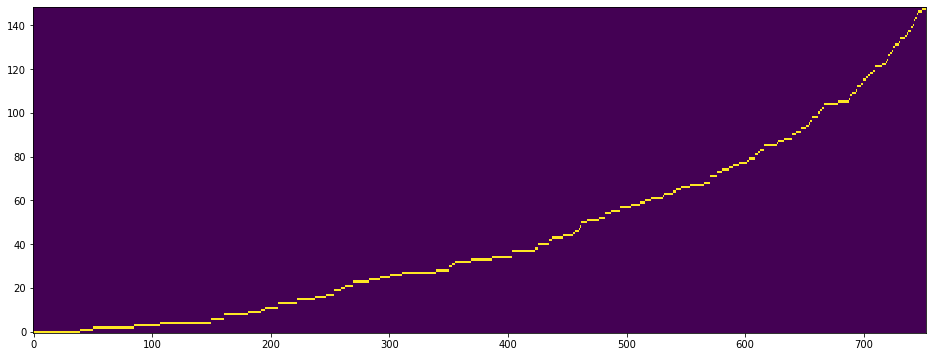}
        %\caption{Processed alignment}
        %\label{fig:processed_alignment}
    \end{subfigure}    
    \caption{Left: source alignment. Right: processed alignment.}
    \label{fig:rap_speedalignment}
\end{figure}

We show examples where we transfer the rhythm from an excerpt by Nicki Minaj to Sally. We showcase the procedural capabilities of Mellotron by processing the source rhythm with a function that produces an \textit{accelerando} starting at half the speed and accelerating to twice the speed. For comparison, we also provide samples conditioned on the pitch contour from Nicki's track. Figure \ref{fig:rap_speedalignment} shows the alignment maps.

\label{sec:audio_rhythmpitch_transfer}
\subsubsection{Rhythm and Pitch Transfer}
% I AM AI -> Sally
% You have a voice that you like but not the rhythm
% Combining rhythm from one speaker and then f0 and voice from training speaker
By conditioning on both rhythm and pitch, we can express characteristics of the source speaker's style. An interesting application is the creation of a hybrid with the style from a source speaker but the voice from another speaker. We show an example where we transfer the characteristics of a solemn speech to Sally. We see that Mellotron contains the same pauses and speech rate as the source which adds to the solemnity of the speech. For comparison, we provide the same phrases synthesised with the original Tacotron 2 which fails to convey the same solemnity. 

\subsection{Singing Voice Synthesis}
Mellotron is able to generalize to rhythm and pitch from styles and speakers not in the training set. We are able to synthesize singing voice from a wide range of input speakers across a range of music styles such as rap, pop, Hindustani and western European classical music. 

\subsubsection{Singing Voice from Audio Signal}
\begin{figure}[!ht]
    \centering
    \includegraphics[width=\linewidth]{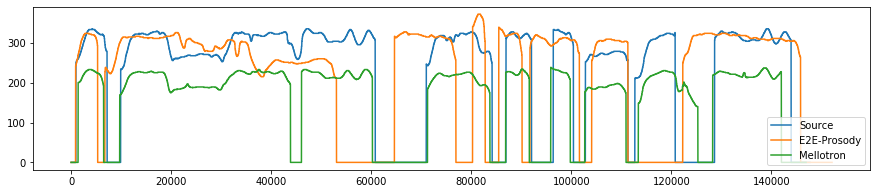}
    \caption{Source, Mellotron and E2E-Prosody pitch contours.}
    \label{fig:singing_sweetdreams}
\end{figure}

Figure~\ref{fig:singing_sweetdreams} shows an example where we use the \textit{Sweet Dreams} sample from the E2E-Prosody paper \cite{skerry2018towards} and transfer its text, rhythm and scaled pitch to Sally. Figure~\ref{fig:singing_sweetdreams} shows that Mellotron's pitch contour is closer to the source than E2E-Prosody is.

\subsubsection{Style transfer from Music Score}
Unlike the experiments on audio, the rhythm and pitch contours provided to the model by a music score are correct by design. We provide a 4-part  example with 20 voices per part on an excerpt of Handel's \textit{Hallelujah}, a 8-part example with 1 voice per part on Ligeti's \textit{Lux Aeterna} and a single voice example synthesizing the opening flute intro from Debussy's \textit{Prélude à l'après-midi d'un faune}. Except from cases where the pitch is beyond the speaker's vocal range, such as in Handel's sample, Mellotron has very precise pitch and rhythm.
\section{Conclusion}\label{sec:discussion}
%auto-ignore
In this paper we described Mellotron, a multispeaker voice synthesis model that allows for direct control of style by conditioning on rhythm and pitch obtained from an audio signal or a music score. 

Our numerical results show that Mellotron is superior to other models with respect to F0 Frame Error. Our qualitative results show that Mellotron is able to generate speech in a variety of styles ranging from read speech to expressive speech, from slow drawls to rap, and from monotonous voice to singing voice although none of these styles are present in the training data. 

Recent singing voice synthesis papers \cite{lee2019adversarially} state that "even in the case of a real recording sample recorded by listening to the original midi accompaniment, it is not easy to adjust the timing and pitch of the correct note" indicating that it is difficult for professional human singers and synthesized voice to match a source audio or source music score perfectly. Our results show that one of the advantages of Mellotron is that the rhythm and pitch contour of a synthesized sample is extremely similar to the source audio file or music score, under the assumption that the pitch is within a speaker's vocal range. When outside a speaker's vocal range, Mellotron defaults to either the lowest tone or highest tone.

For future work, we plan to study the effect of rhythm and pitch contours on the audio quality by comparing samples conditioned on pitch and rhythm data obtained from audio signals versus music scores. With respect to pitch, we are also interested in understanding the effect of multi-speaker training on a speaker's vocal range and extending a speaker's vocal range as much as possible. Last, we would like to train Mellotron on a animated and emotive storytelling style dataset to investigate the contribution of such dataset to Mellotron.
\newpage
\bibliographystyle{IEEEbib}
\bibliography{references}
\end{document}